\documentclass[epj]{svjour}
\usepackage{latexsym,graphicx,epsfig,psfrag,here,color,citesort}    
\def\lesssim{\mathrel{\hbox{\rlap{\hbox{\lower4pt\hbox{$\sim$}}}\hbox{$<$}}}}
\def\query#1{\marginpar{\begin{flushleft}\footnotesize#1\end{flushleft}}}

\begin{document}

\def\eq{\begin{eqnarray}}
\def\en{\end{eqnarray}}

\def\query#1{\marginpar{\begin{flushleft}\footnotesize#1\end{flushleft}}}%

\title{Kaon-nucleon scattering lengths from kaonic 
deuterium experiments}
\author{Ulf-G. Mei{\ss}ner\inst{1,2} \and Udit Raha\inst{1} \and 
 Akaki Rusetsky\inst{1,3}}
\institute{
Universit\"{a}t Bonn, Helmholtz-Institut f\"{u}r Strahlen- und 
Kernphysik (Theorie),\\ 
Nu\ss allee 14-16, D-53115 Bonn, Germany
\and 
Forschungszentrum J\"{u}lich, Institut f\" {u}r Kenphysik (Theorie),
D-52425 J\"{u}lich, Germany
\and
On leave of absence from: High Energy Physics Institute,
Tbilisi State University,\\
University St.~9, 380086 Tbilisi, Georgia
}
\date{Received: date / Revised version: date}
\abstract{
The extraction of the S-wave kaon--nucleon scattering lengths $a_0$ and 
$a_1$ from a  combined analysis of existing kaonic hydrogen and 
synthetic deuterium data has been 
carried out within the framework of a low-energy effective field theory. 
It turns out that with the present  DEAR central values for the kaonic 
hydrogen ground-state energy and width, a solution for $a_0$ and $a_1$ exists 
only in a restricted domain of input values for the kaon-deuteron scattering 
length. Consequently, measuring this scattering length  imposes
stringent constraints on the theoretical description of the kaon-deuteron
interactions at low energies.
}

\maketitle

\section{Introduction}
\label{sec:intro}

Recently, the DEAR collaboration at LNF-INFN has performed a measurement
of the energy level shift and width of the kaonic hydrogen ground 
state~\cite{Beer} with a considerably better accuracy than the earlier
KpX experiment at KEK~\cite{KEK}. The preliminary result of DEAR is
\eq
\epsilon_{1s}&=&193\pm 37~\mbox{(stat)}\pm 6~\mbox{(syst) eV}\, ,
\nonumber\\[2mm]
\Gamma_{1s}&=&249\pm 111~\mbox{(stat)}\pm 30~\mbox{(syst) eV}\, .
\en
As can be  seen from the above result, the accuracy is still tens of eV in the
energy shift and more than 100 eV in the width. Now DEAR is being followed
by the SIDDHARTA experiment that will feature new silicon drift detectors.
The plans of the SIDDHARTA collaboration include the measurement
of both the energy shift and the
width of kaonic hydrogen with a precision of several
eV, i.e. at the few percent level, by 2007. Moreover, SIDDHARTA will attempt the
first ever measurement of the energy shift of the kaonic deuterium 
with a comparable accuracy and
possibly, kaonic helium and sigmonic atoms.

The necessity to perform  measurements of the kaonic deuterium ground-state
observables is justified by the fact that, unlike in the case of  pionic
atoms, the measurement of only the kaonic hydrogen spectrum does not allow
-- even in principle -- 
to extract independently both S-wave $\bar KN$ scattering lengths $a_0$ and 
$a_1$. This happens because the imaginary parts of these scattering lengths
do not vanish in the isospin limit, being determined by the decays 
into inelastic strong channels $\pi\Sigma,\pi^0\Lambda,\cdots$. Consequently, 
one attempts here to
determine four independent quantities (real and imaginary parts of $a_0$
and $a_1$) that requires performing four independent measurements -- e.g., the 
energy level shifts and widths of  kaonic hydrogen {\em and} 
kaonic deuterium.
However, even though it is clear that $a_0$ and $a_1$ can not be determined 
separately without measuring kaonic deuteron, it is still not evident
whether it is possible to do so if one performs such a measurement.
The reason is that the (complex) kaon-deuteron amplitude at threshold, which
is directly determined from the experiment and which is expressed in terms of
$a_0$ and $a_1$ through the multiple-scattering series, is generally
plagued by systematic uncertainties due to a
poor knowledge of the low-energy kaon-nucleon dynamics.
Thus, we have to understand in advance, whether these uncertainties are small 
enough not to hinder a determination of $a_0$ and $a_1$
from the forthcoming SIDDHARTA experiment. This is
the main purpose of the present paper.

The kaon-deuteron scattering process has been extensively studied in the past
within potential (multiple scattering) approaches (see, e.g.
\cite{Hetherington,Torres:1986mr,Deloff:1999gc,Barrett:1999cw}).
To our knowledge, up to now the most detailed investigation of the problem
within an effective field theory (EFT) framework is carried out in 
Ref.~\cite{Kamalov:2000iy} on the basis of the so-called chiral unitary approach. 
In the present paper we undertake a new investigation of the problem
within a systematic field-theoretical approach which, at the lowest
order, is formally similar to that of Ref.~\cite{Kamalov:2000iy}. 
In addition to the previous studies, in this paper
we also consider different sources of the 
systematic error in the calculations, and try to give -- whenever possible -- 
a crude estimate of the theoretical uncertainty.

The main difference of the present article to previous work is, however,
that the existing approaches were exclusively concentrated on the prediction
of the $K^-d$ scattering length from the input $\bar KN$ scattering lengths.
We are not aware of the ``reversed'' analysis in the literature, where the
$\bar KN$ scattering lengths are determined from the input data of kaonic
hydrogen and deuterium ground-state shift and width.
However, this is exactly the type of the analysis that
will be required in the near future 
for the SIDDHARTA data. In this paper -- in the 
absence of any experimental data -- we use ``synthetic'' data and 
show that the reversed calculations, owing to the non-linear 
dependence of the kaon-deuteron amplitude on the $\bar KN$ scattering 
lengths, turn out to be much more sensitive to the theoretical input on
the deuteron structure and the kaon-deuteron interactions, 
than a straightforward evaluation of the $K^-d$ scattering
length through the multiple-scattering series. This fact could potentially
render a combined analysis of the hydrogen and deuterium data a beautiful
testing ground for different EFT descriptions of the 
low-energy kaon-deuteron interactions and, 
as a result, might enable one to accurately
determine the values of the scattering lengths $a_0$ and $a_1$.

The paper is organized as follows. In section~\ref{sec:deuterium} we 
consider the extraction of the complex kaon-deuteron scattering length from the
data on kaonic deuterium. The multiple-scattering series for the kaon-deuteron
scattering amplitude is studied in section~\ref{sec:multiple} and the issue 
of isospin breaking is addressed in section~\ref{sec:isospin}. Further, in 
section~\ref{sec:numerics} the numerical results of the simultaneous analysis
of the existing kaonic hydrogen and synthetic kaonic deuterium data are 
presented and discussed. Section~\ref{sec:concl} contains our conclusions.

\section{Kaonic deuterium}
\label{sec:deuterium}

In the experiments on hadronic atoms one measures the energy levels and
widths of this sort of bound states. At present, there exists a well 
established systematic procedure for extracting 
the values of the pertinent hadronic 
scattering amplitudes at threshold from these measurements, based on
non-relativistic effective Lagrangians (see e.g. 
\cite{pipi,Bern,piK,piN,Raha1,Raha2}). In particular, the case
of the kaonic hydrogen has been addressed in Ref.~\cite{Raha1}.
The calculation of the ground state energy of  kaonic deuteron is 
completely analogous to the derivation of the Deser-type formula
 in the case of pionic deuterium, which was carried
out in Ref.~\cite{Raha2} -- we do not repeat this derivation here. 
However, as first pointed out in Ref.~\cite{gasserK}, in the case of kaonic
atoms one should in addition check whether the relatively large decay widths
of these atoms leads to a
conflict with the use of the Rayleigh-Schr\"odinger perturbation expansion
for the energy levels. In that article it is proposed to estimate the
quantity $\tau E_{1s}^d$, where $\tau$ denotes the lifetime and
 $E_{1s}^d$ the binding energy
for the ground state. Should it turn out that this product is of order one,
using Rayleigh-Schr\"odinger perturbation theory would be questionable.

The lifetime of  kaonic deuterium is determined by the imaginary part
of the $K^-d$ scattering length. In the absence of any reliable experimental 
information, we use the crude estimate 
${\rm Im}\, A_{Kd}\simeq 1~{\rm fm}$ which is consistent with the analysis of
Ref.~\cite{Sibirtsev}.
The decay width and the ground-state binding energy at lowest order in the 
fine-structure constant $\alpha$ are given by
\eq
&&\Gamma_{1s}^d=\frac{1}{\tau}\simeq 4\alpha^3\mu_r^2\,{\rm Im}\, A_{Kd}
\simeq 1.2~{\rm keV}\, ,
\nonumber\\[2mm]
&&E_{1s}^d\simeq\frac{1}{2}\,\mu_r\alpha^2\simeq 10.4~{\rm keV}~.
\en
This yields $\tau E_{1s}^d\simeq 8.6$, which is still large enough to justify 
using perturbation theory (here, $\mu_r$ denotes the reduced mass of the
kaon-deuteron system). 

The (complex) kaon-de\-u\-te\-ron
scattering length $A_{Kd}$ can be extracted from the future SIDDHARTA
data by using
the Deser-type formula at next-to-leading order in isospin breaking, which is
the same as in Ref.~\cite{Raha2}
\eq\label{eq:Deser}
\epsilon_{1s}^d&-&i\,\frac{\Gamma_{1s}^d}{2}=-2\alpha^3\mu_r^2\,
A_{Kd}
\nonumber\\[2mm]
&\times&\bigl\{1-2\alpha\mu_r\,A_{Kd}\,(\ln\alpha-1)
+\cdots\bigr\}\, ,
\en
where the ellipses stand for small contributions which
can be neglected at the accuracy we are working. On the other hand,
since  $\tau E_{1s}^d$ is not very large, it is not excluded
that the corrections at next-to-next-to-leading order in isospin breaking
(which are expected to amount to a few percent)
should also be taken into account, when the accuracy of SIDDHARTA data
is close to the planned one. It is however, well-known that such calculations 
can be performed in a straightforward manner within the non-relativistic 
EFT and hence,  no uncontrollable systematic error finally
emerges at this place. One may therefore safely assume
that the quantity $A_{Kd}$ is directly determined from the 
experiment.

\section{Multiple-scattering series for the kaon-deuteron amplitude}
\label{sec:multiple}

The main purpose of the present paper is to investigate whether 
the measurement of the quantity $A_{Kd}$, along with the $K^-p$ elastic 
scattering amplitude at threshold that is separately 
determined in the experiment 
on kaonic hydrogen, enables one to extract precise values of $a_0$ and $a_1$.
The first step is to express $A_{Kd}$ in terms of $a_0$ and $a_1$ (and
possibly, other physical parameters characterizing the low-energy $\bar KN$
interaction) within the multiple-scattering theory. This procedure can
be appropriately formulated by using the language of effective 
non-relativistic Lagrangians (see, e.g.~\cite{Borasoy,Beane,Raha2}) that,
in addition, allows one to systematically calculate the corrections. 
However, a strong predictive power of the non-relativistic
approach is in fact based on a very subtle balance of different momentum
scales involved in the problem~\cite{Bernard,Raha2}. 
Namely, in this approach the couplings of the effective
Lagrangian are easily expressed in terms of the threshold parameters of the
elementary $\bar KN$ interactions and 
the perturbative expansion of the kaon-deuteron scattering amplitude
corresponds to the usual multiple-scattering series. 
In analogy with the case of the pion-deuteron system one would, however, 
expect that the three-body low-energy constant (LEC), which describes contact 
interaction between the kaon and two nucleons, is strongly enhanced and,
since the value of this LEC is unknown, this leads to a large systematic
error in the predicted kaon-deuteron scattering 
length~\cite{Beane,Borasoy,Raha2}. On the other hand,
if one uses the EFT with non-perturbative pions 
based on the Weinberg counting rules~(see \cite{Epelbaum} for a recent review),
 both the dimensional analysis and resonance saturation~\cite{Raha2},
as well as the study of the scale dependence~\cite{Bernard,HNogga}
carried out in 
the pion-deuteron case, indicate on a rather small uncertainty due to the 
three-body interactions -- 
at the cost of the fact that now the expansion of the pion-deuteron
scattering length should be
carried out in Chiral Perturbation Theory (ChPT) and not
in terms of the threshold parameters of the underlying $\pi N$ scattering
amplitudes. The latter property may lead to even more serious problems in the
kaon-deuteron case, owing to the non-perturbative character of the low-energy
$\bar KN$ interactions and the necessity of using the chiral unitary approach
(see, e.g.~\cite{OR,MO,BNW,OPV}).
The solution to the above dilemma lies in the observation of a certain 
hierarchy of various contributions in the theory with non-perturbative
pions, which could be described by the so-called 
modified power counting~\cite{Bernard}.
The existence of such a hierarchy, which can be traced back to the suppression
of the diagrams with the creation/annihilation of  virtual pions,
indicates that the structure of the theory closely resembles that of the
non-relativistic theory up to the three-body contact terms, whose value
has to be determined from the matching of these two theories. Further,
from the direct 
comparison of the expressions of the pion-deuteron scattering lengths
calculated within these two approaches one may conclude that
the lowest-order three-body LEC in the non-relativistic theory can be
 effectively 
omitted, if the deuteron wave function, calculated in the 
EFT with non-perturbative
pions, is used to evaluate matrix elements in the non-relativistic EFT.
Below, we shall use this simple prescription for estimating the value of the 
three-body LEC in the case of the kaon-deuteron system as well.

The pion-deuteron and kaon-deuteron systems differ in one crucial aspect.
As it is well known (see, e.g.~\cite{Kamalov:2000iy} and references therein), 
the multiple-scattering series for the kaon-deuteron scattering 
does not converge
and requires a (partial) re-summation. This can be done most simply
by using the so-called Fixed Center Approximation (FCA), in which the nucleons
are considered to be infinitely heavy. The validity of this approximation has
been studied both in the potential scattering theory (see e.g.~\cite{Faldt})
and in the EFT approach~\cite{Hanhart}. The fact that FCA can be a reasonable
approximation even for $M_K/m_p\simeq 0.5$ (see Ref.~\cite{Kamalov:2000iy}
and references therein) is related to the peculiar cancellations 
at second order, which are discussed in Refs.~\cite{Faldt,Hanhart}
(here, $M_K$ and $m_p$ denote the masses of the charged kaon and the proton, 
respectively).
In the second-order calculations in the non-relativistic EFT, which we have 
carried out, it is indeed possible to analytically 
verify the cancellation of the
corrections to the FCA at leading order in $M_K/m_p$
(here, one can use the corresponding analytic expressions from 
Ref.~\cite{Raha2} after the substitution $M_\pi\to M_K$). 
Moreover, it should be 
pointed out that in this theory the ``binding corrections''~\cite{Faldt} 
should be also included in the second-order term and, as a result, the
cancellation occurs in both isospin channels and not just in one as in
Ref.~\cite{Hanhart}. Numerically, utilizing  dimensional regularization, the 
corrections turn out to be of order of $20-30\%$ for the kaon-deuteron 
system and of
order of a few percent in the pion-deuteron system. We expect these numbers
to further decrease if the wave functions calculated in EFT with 
non-perturbative pions are used in the calculations. Note that, to the
best of our knowledge, there exists no proof of cancellations beyond
second order, but from e.g. the comparison to the exact solution of Faddeev
equations~\cite{Deloff:1999gc} (see also the discussion in Ref.~\cite{Kamalov:2000iy})
one may conclude that the numbers quoted above
give a realistic estimate of the theoretical error due to FCA at all 
orders. It should however, be stressed that the EFT approach which is used
in the present paper provides an appropriate tool for a systematic calculation
of the corrections to the FCA, as well as the inclusion of higher-order 
(derivative) interactions. In order to provide a sufficient theoretical 
accuracy for analyzing SIDDHARTA results, such calculations might be 
necessary in the future.

Using the FCA in the non-relativistic EFT and neglecting derivative interactions,
we arrive at the expression for the kaon-deuteron scattering length,
which is formally similar to the one from Ref.~\cite{Kamalov:2000iy}:
\eq\label{eq:final-Kamalov}
\biggl(1+\frac{M_K}{M_d}\biggr)A_{Kd} =\int_0^\infty dr\,(u^2(r)+w^2(r))\,
\hat a_{kd}(r)\, ,
\en
where $M_d$ is the deuteron mass, $u(r)$ and $w(r)$ denote the usual 
$S-$ and $D-$wave components of the deuteron
wave function, which are normalized via the condition 
$\int_0^\infty dr\,(u^2(r)+w^2(r))=1$, and
\eq\label{eq:ratio-Kamalov}
\hat a_{kd}(r)=\frac{\tilde a_p+\tilde a_n
+(2\tilde a_p\tilde a_n-b_x^2)/r-2b_x^2\tilde a_n/r^2}
{1-\tilde a_p\tilde a_n/r^2+b_x^2\tilde a_n/r^3}+\delta \hat a_{kd}
\nonumber\\
\en
with $b_x^2=\tilde a_x^2/(1+\tilde a_u/r)$. Further,
\eq
\biggl(1+\frac{M_K}{m_p}\biggr)a_{p,n,x,u}=\tilde a_{p,n,x,u}\, ,
\en
where $a_{p,n,x,u}$ denote the threshold scattering ampli\-tu\-des
for $K^-p\to K^-p$,  $K^-n\to K^-n$, $K^-p\to \bar K^0n$ and 
$\bar K^0n\to \bar K^0n$, respectively. Finally, the quantity
$\delta \hat a_{kd}$ is proportional to the three-body LEC
$f_0^K$, which is a counterpart
of the quantity $f_0$ introduced in 
Ref.~\cite{Raha2}. As discussed above, one may assume that $f_0^K$ vanishes,
if the deuteron wave functions $u(r)$ and $w(r)$ are those calculated in the
EFT with non-perturbative pions. We have further 
carried out the dimensional estimate
of the resulting uncertainty, which yields a few percent
theoretical error in the kaon-deuteron scattering length. Thus, the 
whole procedure is consistent.

The calculations with the deuteron wave functions of the non-relativistic
EFT, which at the leading order are given by
\eq\label{eq:deuteronwf_HM}
u(r)=\sqrt{2\gamma}\,{\rm e}^{-\gamma r},\quad\quad w(r)=0\, ,\quad\quad
\gamma^2\simeq m_pE_d~,
\en
are more subtle (here $E_d$ denotes the binding energy of the deuteron). 
Namely, expanding Eq.~(\ref{eq:ratio-Kamalov}) in the 
multiple-scattering series corresponding to an expansion in powers
of $1/r$ and integrating this term-by-term, it is immediately seen that
the subsequent terms diverge worse and worse as $r\to 0$. This means that
the pertinent integrals require renormalization and are scale-dependent
(see e.g. Refs.~\cite{Beane,Borasoy,Raha2}). On the other hand, the integral before
such an expansion is well-defined and does not have any scale dependence.
It is however, clear that the uncertainty which is related to the existence of
(large) three-body contact interactions  can not simply disappear as a 
result of the re-summation of the multiple-scattering series. To solve
this puzzle, note that the non-relativistic theory makes sense only for the
momenta $|{\bf p}|\ll \lambda$, with $\lambda$ being of the order of the
pion mass. Hence, in order to effectively mimic 
the effect of the short-range physics,
one could straightforwardly perform  cutoff regularization of the 
multiple-scattering series and study the cutoff dependence of the calculated
kaon-deuteron scattering length after the re-summation of the series (even now
the limit $\lambda\to\infty$ exists). This can be done most easily through
the replacement $1/r\to \bigl\{1-\exp(-\lambda r)\bigr\}/r$ in 
Eq.~(\ref{eq:ratio-Kamalov}). Further, in order to relate the value of 
$\lambda$ to the scale $\mu$ of dimensional regularization, which is usually
used in the non-relativistic EFT, we calculate the mean value of the operator
$1/r$ between the wave functions given by Eq.~(\ref{eq:deuteronwf_HM}) twice:  
once using dimensional regularization with minimal subtraction and once
using cutoff regularization. The comparison of these two results gives
$\lambda = \sqrt{e}\mu\simeq 1.65\,\mu$ (where $e$ denotes the basis of 
natural logarithm), so that the interval 
$100~{\rm MeV}\leq\mu\leq 250~{\rm MeV}$, which was fixed in Ref.~\cite{Raha2},
is mapped onto the interval $165~{\rm MeV}\leq\lambda\leq 412~{\rm MeV}$.
Note also that, if instead of the wave functions Eq.~(\ref{eq:deuteronwf_HM})
the wave functions from the EFT with non-perturbative pions are used,
the calculated kaon-deuteron scattering length has practically no
$\lambda$-dependence for $165~{\rm MeV}\leq\lambda\leq \infty$. This shows
that it is consistent to use an ``unregularized''  $1/r$ operator together
with these wave functions, because the necessary cutoff is provided by 
the wave function itself.

\section{Isospin breaking}
\label{sec:isospin}

The equation~(\ref{eq:ratio-Kamalov}) contains four different 
combinations of the threshold amplitudes. Consequently, one has first
to relate these amplitudes to the two scattering lengths $a_0$ and
$a_1$, which should then be determined from the analysis of the
combined data on  kaonic hydrogen and deuterium. 
In this work we take into account the leading-order isospin-breaking
corrections in the kaon-nucleon scattering 
amplitudes which are due to the unitary cusps~\cite{Raha1}.
The re-summation of the bubble diagrams leads to the following simple 
parameterization 
\eq\label{eq:cusps}
a_p&=&\frac{\frac{1}{2}\,(a_0+a_1)+q_0a_0a_1}{1+\frac{q_0}{2}\,(a_0+a_1)}\, ,
\quad
a_n\!=\!a_1\, ,
\nonumber\\
a_x&=&\frac{\frac{1}{2}\,(a_0-a_1)}{1-\frac{iq_c}{2}\,(a_0+a_1)}\, ,
\quad
a_u\!=\!\frac{\frac{1}{2}\,(a_0+a_1)-iq_ca_0a_1}{1-\frac{iq_c}{2}\,(a_0+a_1)}\, ,
\nonumber\\
&&
\en
where
\eq
&&q_c=\sqrt{2\mu_c\Delta}\, ,\quad
q_0=\sqrt{2\mu_0\Delta}\, ,
\nonumber\\[2mm]
&&\Delta=m_n+M_{\bar K^0}-m_p-M_K\, ,
\nonumber\\[2mm]
&&\mu_c=\frac{m_pM_K}{m_p+M_K}\, ,\quad
\mu_0=\frac{m_nM_{\bar K^0}}{m_n+M_{\bar K^0}}\, ,
\en
with $m_n$, $M_{\bar K^0}$ being the masses of the neutron and the 
$\bar K^0$, respectively. Further,
in the calculations we have used the input scattering lengths evaluated
within various versions of the so-called chiral unitary 
approach~\cite{MO,OPV,BNW}, as well as the
experimental values from Ref.~\cite{Martin}. Table~\ref{tab:input} collects
these input values. Note that for us it is not possible to straightforwardly 
use the scattering lengths
(isospin basis) from Refs.~\cite{OR,Kamalov:2000iy}, since the definition
of these scattering lengths (in the isospin limit)
differs from the one, which is used
here or in Refs.~\cite{MO,OPV,BNW}. In particular, in 
Refs.~\cite{OR,Kamalov:2000iy} $a_0,a_1$ are determined from the 
$\bar KN$ scattering amplitude with CM energy taken at $K^-p$ threshold, 
whereas the masses of all particles in the isospin limit are taken
equal to the average masses in the isospin multiplets\footnote{We are 
thankful to E. Oset and J. Oller for a clarifying discussion on this 
and other related topics.}.
Note that, in contrary to this, the usual definition of the scattering 
lengths in the isospin limit
of QCD plus QED implies that the CM energy is set exactly equal to the threshold
energy.

\renewcommand{\arraystretch}{1.6}
\begin{table}[t]
\begin{center}
\caption{$\bar KN$ scattering lengths $a_0$ and $a_1$ (in fm) from 
the literature. 
These scattering lengths are used as an input in the calculations of the
kaon-deuteron scattering length.}
\label{tab:input}
\begin{tabular}{|l|l|l|}
\hline
Ref. & $a_0$ & $a_1$\\
\hline
Mei\ss ner and Oller \cite{MO} & $-1.31 + i 1.24 $ & 
$0.26 + i 0.66  $ \\
\hline
Borasoy {\em et al.}, fit u \cite{BNW} & $-1.48 + i 0.86 $ &
 $0.57 + i 0.83 $ \\
\hline
Oller {\em et al.}, fit A4 \cite{OPV} & $-1.23 + i 0.45 $ &
 $0.98 + i 0.35 $ \\
\hline
Martin \cite{Martin} & $-1.70 + i 0.68 $ &
$0.37 + i 0.60 $ \\
\hline
\end{tabular}
\end{center}
\end{table}

\section{Numerical results and discussion}
\label{sec:numerics}

The tables~\ref{tab:direct_HM}, \ref{tab:direct_soft} and \ref{tab:isospin}
contain our results of the calculations of the kaon-deuteron scattering length 
with the use of the above formulae. In particular, in table~\ref{tab:direct_HM}
the results obtained by using the leading-order deuteron wave function in the
non-relativistic EFT (see Eq.~(\ref{eq:deuteronwf_HM})) are displayed 
for different
values of the cutoff parameter $\lambda$ (or, equivalently, $\mu$). 
Table~\ref{tab:direct_soft} contains the results obtained by using the Paris
and Bonn potential model wave functions as well as the deuteron wave functions
calculated at NLO in the EFT with non-perturbative pions for two different
values of the cutoff parameter $\Lambda$, which is present in this 
theory.
From  table~\ref{tab:direct_HM} one sees -- 
as expected from Refs.~\cite{Beane,Borasoy,Raha2} --  a rather strong
dependence of the calculated kaon-deuteron scattering length on the cutoff
parameter $\lambda$, which has not disappeared after carrying out the 
re-summation of the multiple-scattering series. 
On the other hand, the dependence on the parameter 
$\Lambda$ in table~\ref{tab:direct_soft} is very weak 
(cf. with~\cite{Bernard,HNogga}) and all wave functions in this table yield
practically the same result, which in most cases -- except for the input from 
Ref.~\cite{OPV} -- coincides with the result of
table~\ref{tab:direct_HM} somewhere around $\mu=M_\pi$. Consequently, 
this yields an useful ``rule of the thumb'' for a rough estimate of the value 
of the 
three-body LEC $f_0^K$ in the non-relativistic EFT.

\begin{table*}[t]
\begin{center}
\caption{Kaon-deuteron scattering length $A_{Kd}$ (in fm), calculated by
using Eqs.~(\ref{eq:final-Kamalov},\ref{eq:ratio-Kamalov}) (setting 
$\delta \hat a_{Kd}=0$) and the input from table~\ref{tab:input}. 
The deuteron is described by the wave function in Eq.~(\ref{eq:deuteronwf_HM}). The 
values of the regularization parameter $\lambda$ shown in this table,
correspond to the following values of the dimensional regularization scale:
$\mu=100~{\rm MeV}$; $\mu=M_\pi$; $\mu=250~{\rm MeV}$; $\mu=\infty$ 
(cf. with Ref.~\cite{Raha2}).}
\label{tab:direct_HM}
\noindent\begin{tabular}{|c|c|c|c|c|}
\hline
Ref. &$\lambda=165~\mbox{MeV}$&$\lambda=231~{\rm MeV}$
&$\lambda=412~\mbox{MeV}$&$\lambda=\infty$\\
\hline
\cite{MO}     
       &$-1.35+i\,1.58$&$-1.31+i\,1.42$&$-1.21+i\,1.24$&$-1.07+i\,1.13$\\
\hline
\cite{BNW}    
       &$-1.63+i\,1.39$&$-1.56+i\,1.24$&$-1.41+i\,1.08$&$-1.25+i\,0.99$\\
\hline
\cite{OPV}    
       &$-1.30+i\,1.08$&$-1.30+i\,0.96$&$-1.18+i\,0.82$&$-1.01+i\,0.75$\\
\hline
\cite{Martin} 
       &$-1.70+i\,1.26$&$-1.61+i\,1.14$&$-1.46+i\,0.99$&$-1.31+i\,0.90$\\
\hline
\end{tabular}
\end{center}
\end{table*}

\begin{table*}[t]
\begin{center}
\caption{The same as in table~\ref{tab:direct_HM}, but with the deuteron wave
functions calculated in the Paris and Bonn potential models as well as
 at NLO in EFT with
non-perturbative pions for two different values of the cutoff parameter:
NLO(1) stands for $\Lambda=450~\mbox{MeV}$ and NLO(2) for 
$\Lambda=600~\mbox{MeV}$. The parameter $\lambda$ is set to $\infty$.}
\label{tab:direct_soft}

\noindent\begin{tabular}{|c|c|c|c|c|}
\hline
Ref. & Paris & Bonn & NLO (1) & NLO (2)\\
\hline
\cite{MO} &$-1.29+i\,1.66$&$-1.28+i\,1.65$&$-1.29+i\,1.66$&$-1.30+i\,1.67$\\
\hline
\cite{BNW}
       &$-1.55+i 1.48$&$-1.54+i\,1.47$&$-1.55+i\,1.48$&$-1.56+i\,1.49$\\
\hline
\cite{OPV}
       &$-1.10+i 1.10$&$-1.10+i\,1.09$&$-1.09+i\,1.10$&$-1.11+i\,1.11$\\
\hline
\cite{Martin}
       &$-1.66+i\,1.28$&$-1.65+i\,1.27$&$-1.66+i\,1.28$&$-1.67+i\,1.29$\\
\hline
\end{tabular}
\end{center}
\end{table*}

Finally, in table~\ref{tab:isospin} we show the effect of the isospin breaking
on the theoretical value of the kaon-deuteron scattering length. Bearing
in mind the large isospin-bre\-a\-king corrections in the $\bar KN$ 
amplitudes~\cite{Raha1}
as well as in the pionic deuterium~\cite{Raha3},
it is a little surprising that in most cases -- again except the input from 
Ref.~\cite{OPV} -- the leading-order
isospin-breaking effect at the end turns out to 
be very small. Note also that the large isospin-breaking corrections, which
are quoted in Ref.~\cite{Kamalov:2000iy}, most probably result from the peculiar
definition of the scattering lengths in the isospin limit, which is adopted 
there.

\begin{table*}[t]
\begin{center}
\caption{Isospin-breaking corrections to the $Kd$ scattering length for
different input values of $a_0$ and $a_1$. NLO EFT wave functions are used
for the deuteron. The label ``sym'' refers to the scattering length evaluated
in the isospin symmetry limit.}
\label{tab:isospin}
\noindent\begin{tabular}{|c|c|c|c|c|}
\hline
Ref. &  NLO(1), sym & NLO (1) & NLO(2), sym & NLO(2)\\
\hline
\cite{MO}     
&$-1.26+i\,1.65$
&$-1.29+i\,1.66$
&$-1.27+i\,1.66$ 
&$-1.30+i\,1.67$\\
\hline
\cite{BNW}    
&$-1.48+i\,1.46$
&$-1.55+i\,1.48$
&$-1.50+i\,1.47$ 
&$-1.56+i\,1.49$\\
\hline
\cite{OPV}    
&$-0.85+i\,1.00$
&$-1.09+i\,1.10$
&$-0.87+i\,1.01$ 
&$-1.11+i\,1.11$\\
\hline
\cite{Martin} &$-1.64+i\,1.22$
&$-1.66+i\,1.28$
&$-1.66+i\,1.24$ 
&$-1.67+i\,1.29$\\
\hline
\end{tabular}
\end{center}
\end{table*}

\begin{figure}[t]
\begin{center}
\vspace*{.7cm}
\includegraphics[width=8.cm]{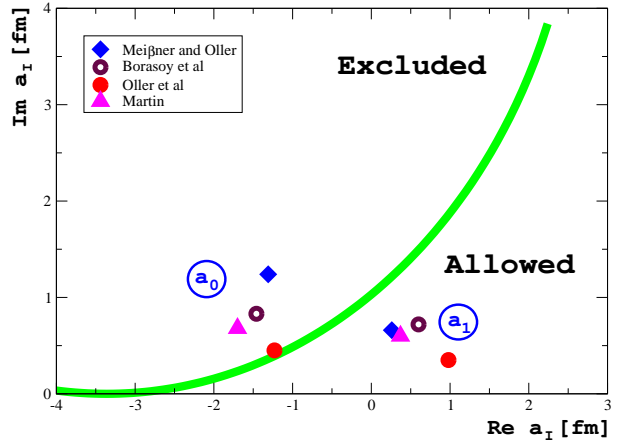}
\end{center}
\caption{Restrictions set by the DEAR data on the values of the scattering 
lengths $a_0$ and $a_1$. For comparison, we give the scattering length 
calculations from different analyses: 
1) Mei{\ss}ner and Oller~\cite{MO}, 
2) Borasoy, Ni{\ss}ler and Weise, fit u~\cite{BNW},
3) Oller, Prades and Verbeni, fit A4~\cite{OPV},
4) Martin~\cite{Martin}.
}
\label{fig:circle}
\end{figure}

 Next, we turn to the main goal of the present paper: 
how does one extract 
precise data of the kaon-nucleon scattering lengths $a_0$ and $a_1$
from the combined measurements
of the (complex) ground-state energy in the kaonic hydrogen and kaonic 
deuterium, which are carried out by the DEAR/SIDDHARTA collaboration?
Let us start from the kaonic hydrogen case, for which
the preliminary results of the measurements are already known.
These measurements alone do not suffice to determine the values of both
$a_0$ and $a_1$, but yield a relation between these two quantities that
leads to a strong restriction on the possible values that $a_0$ and $a_1$ can 
take. In order to understand how this restriction emerges, let us consider
the expression for the ground-state energy of the kaonic hydrogen at
next-to-leading order in isospin breaking~\cite{Raha1}
\eq
\epsilon_{1s}-\frac{i}{2}\,\Gamma_{1s}&=&-2\alpha^3\mu_c^2\, a_p
\bigl\{1-2\alpha\mu_c(\ln\alpha-1)\,a_p\bigr\}\, .
\en
With the help of this equation one may directly
extract the quantity $a_p$ from the DEAR data.
Using now Eq.~(\ref{eq:cusps}) to relate the quantities $a_p$ and 
$a_0,a_1$ we obtain
\eq\label{eq:circle}
a_0+a_1+\frac{2q_0}{1-q_0a_p}\,\, a_0a_1-\frac{2a_p}{1-q_0a_p}=0\, .
\en
Together with the requirement ${\rm Im} \,a_I\geq 0$, which stems from
unitarity, Eq.~(\ref{eq:circle}) defines a circle in the
(${\rm Re}\,{a_I}$, ${\rm Im}\,{a_I}$)--plane. Part of this circle is shown
in Fig.~\ref{fig:circle} (note that, bearing in mind the preliminary 
character of the DEAR data \cite{Beer}, we use only central values
in order to illustrate the construction of
the plot and do not provide a full error analysis). 
In order to be consistent with the DEAR data, both $a_0$ and $a_1$ should be
on the right of this universal DEAR circle. For comparison, on the same figure
we plot $a_0$ and $a_1$ from table~\ref{tab:input}. As we see,
in most of the approaches it is rather
problematic to get a value for $a_0$ which is compatible with DEAR.
This kind of analysis may prove useful in the near future,
when the accuracy of
the DEAR is increased that might stir  efforts on the theoretical side,
aimed at a systematic quantitative description of the $\bar KN$ 
interactions within the unitarized ChPT.

Further, from Eq.~(\ref{eq:circle}) one may determine e.g. $a_1$.
Substituting this expression into Eqs.~(\ref{eq:final-Kamalov}),
(\ref{eq:ratio-Kamalov}) and (\ref{eq:cusps}), one arrives at a 
non-linear equation for determining $a_0$ with a given input value
of $A_{Kd}$.
In the absence of  experimental data on kaonic deuterium, we adopt
the following strategy to solve this equation. In the literature, we find
different theoretical calculations of the $Kd$ scattering length. We consider
some of them, namely the results of 
Refs.~\cite{Torres:1986mr,Deloff:1999gc,Grishina,Bahaoui}, as synthetic 
experimental data and solve the corresponding equation with respect to $a_0$.
For completeness, we also present the results which are obtained 
from the same synthetic deuteron data together with the old
KpX data on kaonic hydrogen, 
$\epsilon_{1s}=323~{\rm eV}$, $\Gamma_{1s}=407~{\rm eV}$
\cite{KEK}. Note also that our  solution includes 
only central values of the input kaon-deuteron and kaon-proton amplitudes. 
At this preliminary stage it is natural to postpone the full error analysis, 
until the SIDDHARTA data on the kaonic deuterium become available.

\begin{table*}[t]
\begin{center}
\caption{The scattering lengths $a_0$ and $a_1$, obtained from
a simultaneous analysis of DEAR (KpX) data and the synthetic input for
the $Kd$ scattering length. The deuteron wave function is described by
Eq.~(\ref{eq:deuteronwf_HM}) and the regularization parameter
$\lambda=132~{\rm MeV}$.}
\label{tab:inverse_HM}
\noindent\begin{tabular}{|l|l|l|}
\hline
\hspace*{1.2cm}Input $A_{Kd}$&DEAR &KpX\\
\hline
$-1.34 + i\,1.04$ \cite{Torres:1986mr}
& $a_0=-1.27 + i\,0.36$ & $a_0=-1.34 + i\,0.63$\\
 & $a_1=1.19 + i\,0.09$  & $a_1=0.01 + i\,0.55$ \\
\hline
$-0.85 + i\,1.10$ \cite{Deloff:1999gc}, Faddeev Eq. 
& $-$ & $a_0=-1.40 + i\,0.82$\\
&& $a_1=-0.09 + i\,0.07$ \\
\hline
$-0.75 + i\,1.12$  \cite{Deloff:1999gc}, FCA 
& $-$ & $a_0=-1.41+ i\,0.84$\\
 &&$a_1=-0.10 + i\,0.01$ \\
\hline
$-0.78 + i\,1.23$ \cite{Grishina}
& $-$ &$a_0=-1.43 + i\,0.82$\\
 &&$a_1=-0.04 + i\,0.01$ \\
\hline
$-1.92 + i\,1.58$ \cite{Bahaoui}
& $-$ & $a_0=-1.64 + i\,0.63$\\
&&$a_1=1.03 + i\,0.02$ \\
\hline
\end{tabular}
\end{center}
\end{table*}

We start with the deuteron wave function in the non-relativistic EFT,
which is given by Eq.~(\ref{eq:deuteronwf_HM}). The results for the extracted
scattering lengths $a_0$ and $a_1$ are given in table \ref{tab:inverse_HM},
where we had to choose 
a rather small value of the cutoff parameter $\lambda=132~{\rm MeV}$
corresponding to $\mu=80~{\rm MeV}$. From this table one already sees the main
property of the solutions: due to a highly non-linear form of the equation
which determines $a_0$ and $a_1$, the solutions do not always exist
(at least for those values of these scattering lengths which can be termed
physically reasonable). Note that this mainly concerns the solutions which
are obtained using DEAR input. The KpX data turn out to be much less 
restrictive. For example, if we increase the value of the parameter $\mu$ even
up to $100~{\rm MeV}$, the sole solution with DEAR input in 
table \ref{tab:inverse_HM} disappears, whereas the solutions with KpX
input are still present. Note also that we have analyzed 
more synthetic input than is finally shown in table \ref{tab:inverse_HM}:
for most of the input values of the kaon-deuteron scattering length which
are known in the literature, there exists no solution for $a_0$ and $a_1$.

Bearing in mind the results in table \ref{tab:inverse_HM}, it is not
surprising that, using the wave functions calculated an NLO in the theory
with non-perturbative pions, the solutions exist only with the KpX input.
The corresponding results are shown in table \ref{tab:inverse_soft}.

\begin{table*}[t]
\begin{center}
\caption{The same as in table \ref{tab:inverse_HM}, but with the NLO EFT 
wave functions and $\lambda=\infty$. Only KpX input is displayed, since
there are no solutions with DEAR input.}
\label{tab:inverse_soft}
\noindent\begin{tabular}{|l|l|l|}
\hline
\hspace*{1.2cm}Input $A_{Kd}$ & $\Lambda=450~\mbox{MeV}$ & $\Lambda=600~\mbox{MeV}$ \\
\hline
$-1.34 + i\,1.04$ \cite{Torres:1986mr}  & 
 $a_0=-1.36+i\,0.59$&$a_0=-1.35+i\,0.60$\\
&$a_1=0.10+i\,0.63$&$a_1=0.07+i\,0.62$\\
\hline
$-0.85 + i\,1.10$ \cite{Deloff:1999gc}, Faddeev Eq.&
 $a_0=-1.41+i\,0.80$&$a_0=-1.41+i\,0.80$\\
&$a_1=-0.04+i\,0.08$&$a_1=-0.05+i\,0.08$ \\
\hline
$-0.75 + i\,1.12$ \cite{Deloff:1999gc}, FCA &
$a_0=-1.42+i\,0.82$&
$a_0=-1.42+i\,0.82$\\
&$a_1=-0.04+i\,0.03$&
$a_1=-0.05+i\,0.03$ \\
\hline
$-0.78 + i\,1.23$ \cite{Grishina} & $a_0=-1.44+i\,0.80$& $a_0=-1.44+i\,0.80$\\
&$a_1=0.01+i\,0.03$ & $a_1=0.00+i\,0.03$\\
\hline
\end{tabular}
\end{center}
\end{table*}

\begin{figure}[t]
\begin{center}
\vspace*{.7cm}
\includegraphics[width=8.cm]{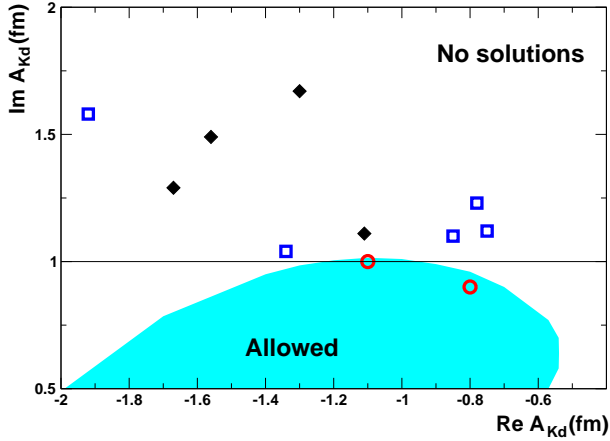}
\end{center}
\caption{The region in the $({\rm Re}\,A_{Kd}$,  ${\rm Im}\,A_{Kd})$--plane
where  solutions for $a_0$ and $a_1$ exists. The cutoff parameter is 
$\Lambda=600~{\rm MeV}$. For comparison, we also show the results of
calculations of the kaon-deuteron scattering length from 
table~\ref{tab:direct_soft} (filled diamonds), the synthetic data from
tables~\ref{tab:inverse_HM}, \ref{tab:inverse_soft} (squares) and
our representative solutions from table~\ref{tab:1.1} (circles). 
}
\label{fig:area}
\end{figure}

One might finally ask the following question: 
how large should the kaon-deuteron scattering 
length be so that a solution for $a_0$ and $a_1$ exists at all? In order
to answer this question, we have scanned the $({\rm Re}\,A_{Kd}$,  
${\rm Im}\,A_{Kd})$--plane in the interval $-2~{\rm fm}< {\rm Re}\,A_{Kd}<0$
and  $0.5~{\rm fm}< {\rm Im}\,A_{Kd}<2.5~{\rm fm}$ 
and tried to find
solutions, using DEAR input data. The results of this investigation, which are
displayed in Fig.~\ref{fig:area}, are very
interesting: it turns out that the solutions exist only 
if ${\rm Im}\,A_{Kd}\lesssim 1~{\rm fm}$ and moreover, if 
${\rm Im}\,A_{Kd}\simeq 1~{\rm fm}$ then one finds solutions only
 in a very small interval around ${\rm Re}\,A_{Kd}\simeq -1~{\rm fm}$. 
Some representative solutions are 
shown in table~\ref{tab:1.1}. We wish to also note that all
this agrees with the scattering data analysis, carried out in 
Ref.~\cite{Sibirtsev}.
 If ${\rm Im}\,A_{Kd}$ crosses the border of the shaded area in 
Fig.~\ref{fig:area} continuously from below, then on the same branch 
one gets the solution with ${\rm Im}\,a_1\leq 0$ that is forbidden 
by unitarity.
On the other hand, if KpX input data 
are used, the solutions exist in a much larger area.

\section{Conclusions}
\label{sec:concl}

 The message of our investigation is very clear.
Up to now, in the theoretical description of the deuteron we have restricted
ourselves to the lowest-order expression in the non-relativistic EFT
(derivative interactions are neg\-lec\-ted) and, in addition, worked in the limit
of  infinitely heavy nucleons (FCA). It is widely believed that these
approximations represent a good starting point for the description of the
kaon-deuteron scattering length. Moreover, in the non-relativistic
EFT the corrections are systematically calculable.

Within the approximations described above it turns out that the 
combined analysis of DEAR/SIDDHARTA
data on kaonic hydrogen and deuterium 
is more restrictive than one would {\it a priori} expect. In particular,
we see that  solutions exist
only in a rather small area of the $({\rm Re}\,A_{Kd}$, ${\rm Im}\,A_{Kd})$--plane.
Due to this fact, in certain cases it might be possible to pin down 
 the values of $a_0$ and $a_1$ at a 
reasonable accuracy, even if $A_{Kd}$ itself is not measured very accurately. 
Moreover, if the corrections to the lowest-order approximate result,
as expected, are moderate, they should not change the qualitative picture.
In our opinion, it could prove very useful to perform the calculations of 
these corrections before starting to analyze the forthcoming SIDDHARTA data 
since, as one may conclude from the discussion in the present paper, 
exactly these corrections constitute the largest potential source of 
theoretical uncertainty at present. On the other hand, crude estimates show
that the uncertainty, coming from the short-range QCD dynamics (three-body LEC)
is rather small and should not likely hinder the analysis of the future 
SIDDHARTA data on the kaonic deuteron.
Note also that the above conclusion concerns DEAR input only. The analysis
using KpX input turns out to be much less restrictive and therefore less
informative than with the DEAR input.

\begin{sloppypar}
To summarize, one may expect that the combined analysis of the forthcoming 
high-precision data from DEAR/SID\-DHAR\-TA collaboration on kaonic hydrogen 
and de\-u\-te\-ri\-um will enable one to perform a stringent test of the 
framework used to describe low--energy kaon-deuteron scattering,
as well as to extract the values of $a_0$ and $a_1$ with a reasonable accuracy.
However, in order to do so, much theoretical work related to the systematic
calculation of higher-order corrections within the non-relativistic EFT 
is still to be carried out.
\end{sloppypar}

\bigskip

\begin{table}[t]
\begin{center}
\caption{Representative solutions for $a_0$ and $a_1$ using the
NLO EFT wave functions
together with DEAR input and ${\rm Re}\, A_{Kd}\simeq -1~{\rm fm}$,
${\rm Im}\, A_{Kd}\simeq 1~{\rm fm}$.}
\label{tab:1.1}
\noindent\begin{tabular}{|l|l|l|}
\hline
\hspace*{.8cm} $A_{Kd}$ &$\Lambda=450~\mbox{MeV}$ &$\Lambda=600~\mbox{MeV}$ \\
\hline
$-1.10 + i\,1.00$ &$a_0=-1.25+i\,0.38$
&$a_0=-1.24+i\,0.38$ \\
& $a_1=1.06+i\,0.00$
&$a_1=1.02+i\,0.03 $\\
\hline
$-0.80 + i\,0.90$ &$a_0=-1.07+i\,0.43$&
$a_0=-1.06+i\,0.42$\\
&$a_1=0.44+i\,0.07$&$a_1=0.42+i\,0.08$ \\
\hline
\end{tabular}
\end{center}
\end{table}

\begin{sloppypar}
{\it Acknowledgments:}
The authors would like to thank J\"urg Gasser, Christoph Hanhart, 
Vadim Lensky, Andreas Nogga and Sasha Sibirtsev for discussions.
We in addition thank Evgeny Epelbaum for providing the values of the NLO 
wave function of the deuteron and Robin Ni{\ss}ler for communicating 
the separate values of $a_0$ and $a_1$ in their approach to us.
Partial financial support from the EU Integrated Infrastructure
Initiative Hadron Physics Project (contract number RII3-CT-2004-506078)
and DFG (SFB/TR 16, ``Subnuclear Structure of Matter'') is gratefully
acknowledged.
\end{sloppypar}

\end{document}